# Continuous melting and thermal-history-dependent freezing in the confined Na-K eutectic alloy

E. V. Charnaya[1,2,*], M. K. Lee[1,3], Cheng Tien[1], L. J. Chang[1], Z.-J. Wu[1], Yu. A. Kumzerov[4], A. S. Bugaev[5]

[1]*Department of Physics, National Cheng Kung University, Tainan, 70101 Taiwan*
[2]*Institute of Physics, St. Petersburg State University, St. Petersburg, 198504 Russia*
[3]*NSC Intrument Center At NCKU, Tainan, 70101 Taiwan*
[4]*A. F. Ioffe Physico-Technical Institute RAS, St. Petersburg, 194021 Russia*
[5]*Moscow Institute of Physics and Technology, Moscow, 141700 Russia*

[23]Na NMR studies of the Na-K eutectic alloy embedded into porous glass with 7 nm pores showed that melting of $Na_2K$ confined nanoparticles is a continuous process with smooth changes in the Knight shift of a narrow resonance line and nuclear spin relaxation between those in the crystalline and liquid states. The intermediate state which occurs upon melting is stable and more favorable than the liquid state. The inverse freezing transformation can be sharp as at a first order transition or continuous depending on the initial temperature of cooling. The results suggest revision of theoretical predictions for the melting and freezing transitions in confined geometry.

Melting in bulk is a pure first-order transition that occurs when the chemical potentials of the solid and liquid states become equal. At a melting temperature the rigid crystalline lattice transforms to the fluid and physical parameters of the matter exhibit abrupt alterations. When the particle size decreases the melting temperature generally decreases due to the stronger impact of surface ([1-3] and references therein). In addition, the formation of a liquid skin on the solid core before melting becomes more important since the liquid layer thickness and the particle size are comparable. When the liquid skin appears, its thickness rises upon further warming until the solid core melts sharply at a particular temperature [4,5]. The inverse process of freezing occurs via nucleation. In small particles there is a large thermal hysteresis between melting and freezing. For ultra-fine clusters the computer simulation predicts a finite temperature range where the solid and liquid states coexist dynamically, clusters remaining solid or liquid below or above this range, respectively [6,7].

Experimental studies of melting for small metallic particles were started long time ago and intensified more recently following the development of nanotechnology ([7-10] and references therein). A general decrease in the melting temperature was observed when particles were not subjected to external disturbance [11,12] such as, for instance, mechanical strains, and could be considered free to some extent. The reduction in the melting temperature was found for ensembles of small particles, isolated as well as embedded into nanoporous matrices, and for single nanoparticles. For ensembles of nanoparticles, the diffused melting was usually observed which is believed to be caused by particle size distribution [2,13,14], however, the emergence of the liquid layer as a precursor was also suggested [15]. For individual particles, the formation of a liquid skin was proved by electron diffraction [16]. However, even when the emergence of a liquid skin on the particle surface led to some broadening of the melting transition, the well defined solid and liquid states coexisted within a temperature range and the melting transition in the solid cores remained sharp. Ultra-small supported and unsupported metallic clusters were studied predominantly by various calorimetric methods (see [1,7,8,17-20] and references therein). Smooth caloric curves were observed upon melting in some experiments, for instance for sodium clusters comprising of 139 atoms [8], which agree with computer models, while sharp changes were found for larger sodium clusters.

Whether the continuous melting could happen in particles under nanoconfinement is still a totally open question while it is crucial for understanding the influence of size reduction on the melting process. Calorimetric methods being extremely useful for phase transition studies cannot answer this question as they provide only averaged information. Another challenging problem is whether the continuous melting, if confirmed, coexists with sharp or continuous freezing. Experimental methods which are sensitive to microstructure are required to investigate such matters. Among them, NMR has a lot of advantages for studying phase transitions in confined geometry [14,15,21-24].

Here we present $^{23}$Na NMR studies of the melting and freezing phase transition in the Na-K eutectic alloy embedded into silica porous glass. This alloy precipitates with formation of the cubic intermetallic compound $Na_2K$. The high symmetry of crystalline $Na_2K$ facilitates the observation of resonance lines at freezing. We will show that the position of the $^{23}$Na NMR line (Knight shift) exhibits continuous change upon warming through the solid to liquid transition until complete melting while it shows a step-like jump upon crystallization. The step-like freezing reproduces only after the alloy was completely melted previously. When cooling is started before the Knight shift upon warming achieves its value in the melt, the freezing is also continuous. In agreement with the behavior of the Knight shift, nuclear spin-lattice relaxation changes gradually upon melting which evidences continuous acceleration in ionic mobility.

Liquid alloy of the composition 66.6 at% K and 33.4 at% Na was embedded into porous glass with pore size of 7 nm at room temperature under high pressure. The composite thus obtained remains stable when kept in the nitrogen atmosphere or in oil contrary to the case of pure Na metal when it is introduced at elevated temperature [25].

Fig.1 shows variations of the Knight shift during two different cooling-warming cycles from room temperature down to 150 K and up to 380 K. The Knight shift varies weakly when the confined alloy remains unfrozen. Near 200 K a signal from frozen $Na_2K$ appears and the liquid and crystalline phases coexist within a narrow temperature range. The difference between the Knight shift in liquid and solid at freezing is more than 80 ppm which is much greater that the line width, so the lines from solid and liquid are well separated as can be seen at 204 K (Fig.2).

The behavior of the Knight shift is totally different upon warming. There is no any sharp change which can be identified with the first order phase transition from solid to liquid. Instead, a smooth continuous increase from the value corresponding to that in the crystalline state is seen until the Knight shift reaches the value corresponding to the melt at about 250 K. Such behavior (sharp freezing and continuous melting) is always reproducible after the sample was warmed above the melting accomplishment.

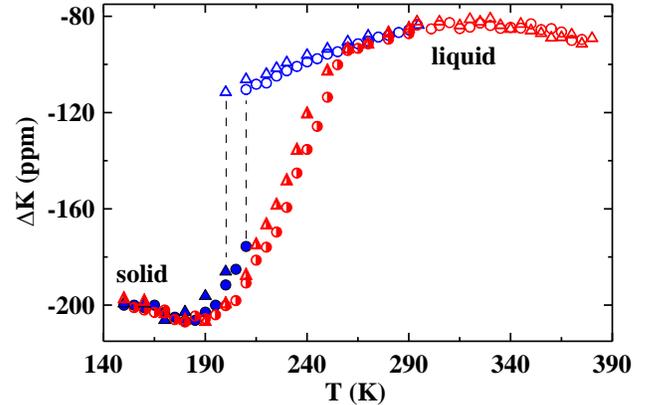

Fig.1. Variations with temperature of the $^{23}$Na Knight shift in the confined Na-K alloy upon cooling from room temperature (open symbols - liquid, closed symbols – solid, blue) and consecutive warming (semi-closed symbols, red) during two thermal cycles marked with circles and triangles. Dashed lines indicate the coexistence of melted and frozen alloy.

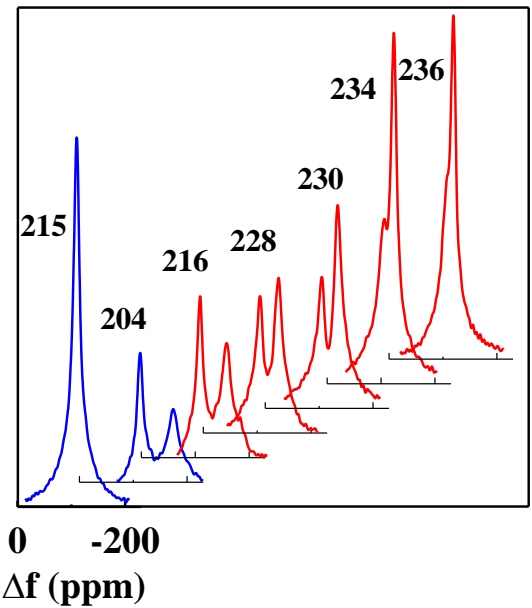

Fig.2. $^{23}$Na NMR spectra upon cooling from room temperature at 215 and 204 K (blue) and subsequent warming (red). Temperatures are indicated in the panel.

If we change the thermal cycling conditions and start warming before the whole amount of the confined liquid alloy is frozen (that is at a temperature where the NMR signals from liquid and crystalline states coexist) two lines can be seen within a rather large temperature range which behavior is quite unexpected (Figs.2 and 3). The line which is associated to the melt decreases gradually in intensity and vanishes completely till 240 K, while a line which emerged as a result of partial freezing and was associated to the signal from crystalline $Na_2K$ increases continuously in intensity and transforms into the signal from melted alloy upon further warming. Such behavior is also fully reproducible.

ascribed to the first order transition from the liquid to solid state. Smooth variations in the Knight shift repeat upon subsequent warming and cooling (Fig.4).

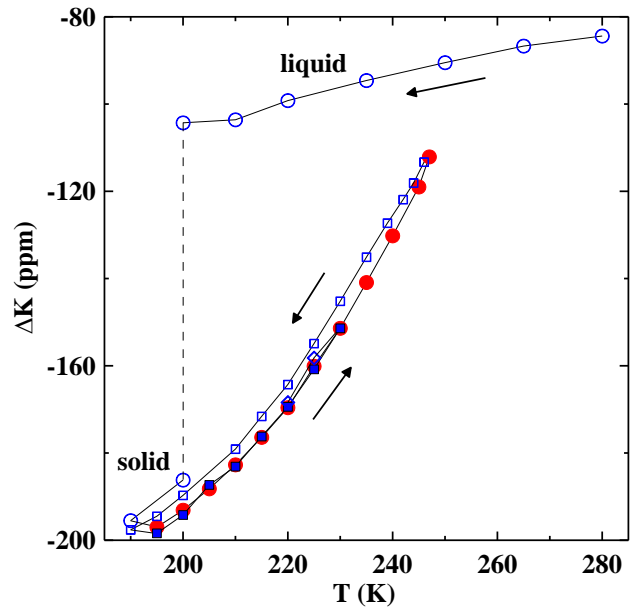

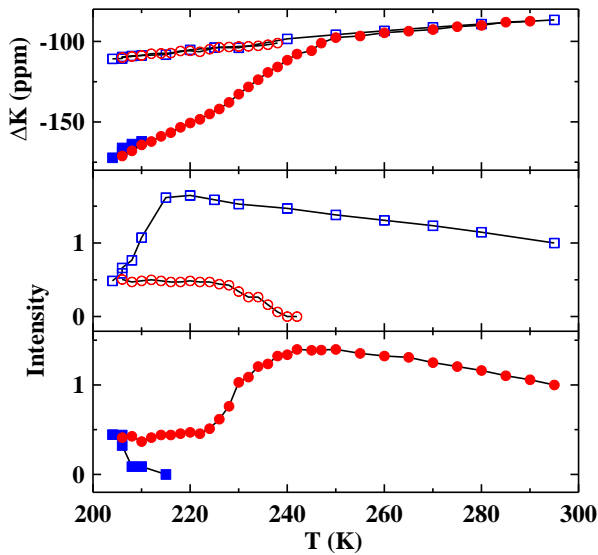

Fig.3. Variations with temperature of the $^{23}$Na Knight shift and intensity of lines upon cooling (squares, blue) and warming (circles, red). Open symbols – NMR line associated to liquid, closed symbols – second component of NMR spectra, associated at low temperature with solid. The integral intensities of the NMR lines are normalized to their values at room temperature. Solid lines are guides for the eye.

When the sample is cooled down till complete freezing in the confined melt, warmed up to a temperature below the accomplishment of melting and then cooling is started again, the freezing phase transition differs drastically from that illustrated by Figs.1-3. The Knight shift changes continuously upon second cooling (Fig.4) until the NMR line coincides with that corresponding to the crystalline state. No any jump can be seen which can be

Fig.4. Variations with temperature of the $^{23}$Na Knight shift upon cooling (open symbols, blue) and warming (closed symbols, red). Circles – cooling from room temperature down to 190K and warming up to 245 K, squares – cooling to 190 K and warming to 230 K, diamonds – cooling to 220 K. Solid curves are guides for the eye. Dashed straight line shows the coexistence of liquid and solid upon first cooling.

The temperature dependences of the NMR spectra, Knight shift and intensity is in accordance with variations of spin-lattice relaxation. As is well known, spin-lattice relaxation of quadrupole nuclei in bulk metals [26,27] goes along two main paths: due to magnetic interaction of the nuclear dipole moments with conduction electrons and due to electric interaction of the nuclear quadrupole moments with dynamic gradients of electric fields resulted from ionic movement. The latter is very sensitive to alterations in mobility under nanoconfinement and becomes more effective for metals embedded into nanopores [22,24]. $^{23}$Na nuclei have spin 3/2 and thus possess the quadrupole moment. Therefore, one can expect that longitudinal spin relaxation for sodium nuclei should change upon melting and freezing in the confined alloy. Fig.5 shows dependences of the $^{23}$Na spin-lattice relaxation time on temperature in the sample under study upon thermal cycling. One can see that the spin-lattice relaxation time changes in the step-like manner when the sample was cooled down from room temperature but it changes continuously upon warming. Upon freezing, relaxation becomes faster in solid. This cannot occur due to alterations in the Knight shift as when the Knight shift decreases, magnetic relaxation slows down according to the Korringa relation [28]. Thus, the effect of ionic mobility dominates relaxation changes at the transition. During partial thermal cycles which correspond to variations in the Knight shift shown in Figs. 3 and 4, longitudinal relaxation changes gradually. Relaxation is faster when ionic mobility slows down and the relevant correlation time becomes closer to the inverse value of the Larmor frequency. The continuous change in spin-lattice relaxation evidences the continuous change in ionic mobility through the transition between the liquid and solid states.

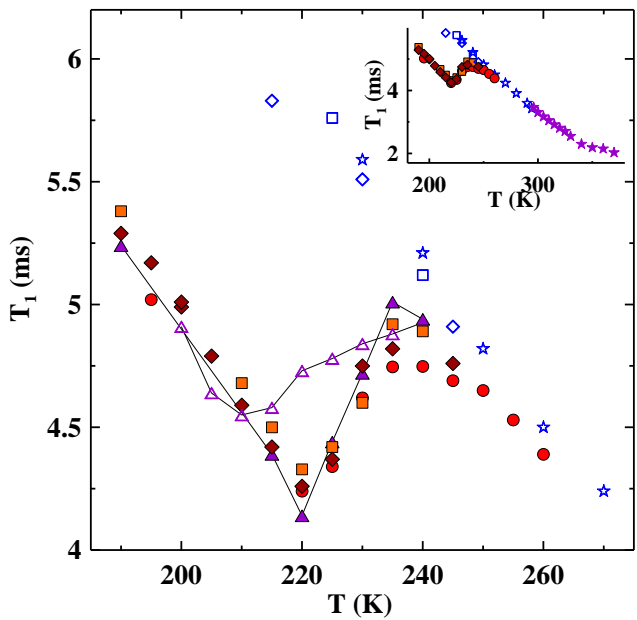

Fig.5. Temperature dependences of the spin-lattice relaxation time for several cooling and warming shown with different symbols. Open (blue) symbols – cooling, closed (various reddish colors) – warming. Triangles correspond to the thermal cycle shown in Fig.4. The solid line is a guide for the eye. The inset shows results in extended temperature range.

The NMR responses which are shown in Figs.1-5 correspond to a big ensemble of nanoparticles embedded into the glass matrix. They are averaged over particles with a size distribution. However, this cannot imitate the observed gradual transformation of the NMR spectrum between the liquid and solid states. The particle size distribution could lead to the broadening of the temperature range where the NMR lines corresponded to solid and liquid coexist but not to smooth alteration in the Knight shift of a narrow NMR line and in the relaxation rate. Note that the whole set of NMR measurements were repeated for two other nanocomposite samples and the results obtained were quite similar. However, melting and freezing were found to be step-like in larger pores of photonic crystals.

The Knight shift is caused by conduction electrons and is proportional to electron susceptibility [28]. In bulk metals the step-like changes in the Knight shift upon melting or freezing agree with alterations in electron density as a result of the first order phase transition. Therefore, the smooth gradual alteration in the Knight shift observed here should be treated as an evidence of continuous structural transformation in the confined alloy upon melting and also upon freezing at particular conditions. Another evidence is given by continuous alteration in spin-lattice relaxation reflecting smooth changes in ionic mobility.

According to the experimental data obtained, melting occurs through formation of an intermediate state. This state shows no noticeable time evolution at a fixed temperature and is more favorable at melting than the liquid state (Fig.3). Freezing which started from this state is continuous. Such experimental results contradict the simplest model of melting in small particles based on the Gibbs-Thompson equation which leads to the melting temperature reduction, only. They apparently do not agree to the liquid skin model as there is no coexistence of the NMR lines from solid core and liquid skin. The continuous changes in the Knight shift and intensity of NMR lines through the melting transition and freezing from the intermediate state could be consistent with the computer models which predict for small clusters the dynamic coexistence of liquid and solid states within a finite temperature range if the NMR spectrum is implied to be merged into a single line with mean parameters due to fast exchange. However, for such a dynamic coexistence model there is no direct way to interpret the gradual changes in spin-lattice relaxation within the temperature range where the intermediate state emerges. There are other experimental results which require additional theoretical elucidation, such as transformation of liquid into the intermediate state upon warming which are shown in Fig.3. The full set of experimental data is in best accordance with the suggestion of smooth softening of the $Na_2K$ structure from the crystalline lattice to liquid while such a model was not discussed until now.

In conclusion, the gradual transformation in the $^{23}Na$ NMR line and spin-lattice relaxation from solid to liquid upon warming evidences continuous melting of confined $Na_2K$ nanoparticles. The melting is irreversible and the freezing is sharp, step-like when cooling is started from completely melted alloy. However, the melting is reversible and the freezing is smooth when cooling starts from the intermediate state.

Authors acknowledge the financial support from National Cheng Kung University (Taiwan) and Russian Foundation for Basic Research.

*Corresponding author: E.V. Charnaya, charnaya@live.com